\documentclass[a4paper,11pt,fleqn]{article}

\usepackage{ucs}
\usepackage[utf8x]{inputenc}
\usepackage{amsmath}
\usepackage{amsfonts}
\usepackage{amssymb}
\usepackage[american]{babel}
\usepackage{array}
\usepackage{epsfig}

 \usepackage{feynmp}
 \usepackage{graphicx}
\newcommand{\Tr}{\text{tr~}}
\newcommand{\tr}{\text{tr~}}

\renewcommand{\title}[1]{\vbox{\center\LARGE{#1}}\vspace{5mm}}
\renewcommand{\author}[1]{\vbox{\center#1}\vspace{5mm}}
\newcommand{\address}[1]{\vbox{\center\em#1}}
\newcommand{\email}[1]{\vbox{\center\tt#1}\vspace{5mm}}
\usepackage{slashed}

\begin{document}
\title{Anomalous Dimensions in Non-Supersymmetric
Bifundamental Chern-Simons Theories}
\author{V. Gurucharan${}^{a}$, Shiroman Prakash${}^{b}$}
\address{Dept. of Physics and Computer Science, Dayalbagh Educational Institute, \\Dayalbagh,
Agra 282005, India.}

\email{$^a$vgurucharan2@gmail.com, $^b$shiroman@gmail.com}

\begin{abstract}
 Non-abelian Chern-Simons theories coupled to fermions are known to provide an interesting class of  non-supersymmetric conformal fixed points \cite{Giombi:2011kc}. These theories, particularly those based on bifundamental matter, are important because they may provide simple non-supersymmetric examples of the AdS/CFT correspondence. For instance, it seems natural to conjecture that $O(N)_{-k}\times O(N)_k$ Chern-Simons theory coupled to Majorana fermions transforming in a bi-vector representation may be dual to pure Einstein gravity with a small negative cosmological constant in the ``M-theory'' limit where $k=1$ and $N$ is large. 
While it is extremely difficult to directly study such bifundamental theories when $k=1$ or even at strong 't Hooft coupling $\lambda=\frac{N}{k}$, it is possible to calculate physical quantities to all orders in $\lambda$ in a $U(M)_{k_M} \times U(N)_{k_N}$ theory, in the limit $M \ll N$, in an $M/N$ expansion. To illustrate this, we calculate the anomalous dimension of the primary operator $\tr \bar{\psi}{\psi}$, to first order in $M/N$, to all orders in $\lambda_M=\frac{N}{k_M}$, but with $\lambda_N=\frac{N}{k_N}=0$. We also comment on possible bosonization dualities for bifundamental Chern-Simons theories.
\end{abstract}

\section{Motivation}
Consider the following theory: $U(N) \times U(N)$ Chern-Simons theory, with
equal and opposite levels, $k$ and $-k$, coupled to a massless
Dirac fermion in the bifundamental representation. As pointed out in \cite{Giombi:2011kc}, this theory is
one of a large number of non-supersymmetric conformal field theories that exist
in 3 dimensions. Though
non-supersymmetric, this theory is clearly conformal (at least in perturbation theory) and, like the well known
supersymmetric $\mathcal N =6$ theory (ABJM)\cite{Aharony:2008ug}, admits a
large $N$ limit. 


The most natural large-$N$ limit is the 't Hooft limit, in which the 't Hooft
coupling $\lambda=\frac{N}{k}$ is held constant as $N \rightarrow \infty$ and $k \rightarrow \infty$. We might consider a slight generalization to two gauge groups with possibly unequal ranks $M$ and $N$ and levels $k_N$ and $k_M$, for each of the two Chern-Simons gauge fields.  

When $M=N$, we can define the  't Hooft couplings $\lambda_+=N/k_N+N/k_M$ and $\lambda_-=N/k_N-N/k_M$; which are both effectively continuous in the 't Hooft limit. If $\lambda_+=0$, then $k_N=-k_M$, and a parity transformation (which flips the signs of $k_N$ and $k_M$) simply interchanges $k_N$ and $k_M$ so the theory is parity invariant. However, when  $\lambda_+ \neq 0$ the theory is not invariant under parity. If we start with the theory with $\lambda_-$ large and $\lambda_+=0$, small changes in $\lambda_\pm$ could be obtained by deforming the theory by appropriate operators; these operators would presumably be dual to dilaton and axion-like fields respectively, in a holographic description \cite{Maldacena:1997re}. 

However, another large $N$ limit may exist, as is well-known since the discovery
of ABJM \cite{Bagger:2007jr, Aharony:2008ug, Benna:2008zy}. (Using supersymmetric localization techniques, it is possible to calculate certain physical quantities exactly in this limit \cite{Marino:2011eh, Grassi:2014vwa}.) ABJM at fixed $k$, say $k=1$, and large $N$ is dual to M-theory on
$AdS_4 \times S_7/Z_k$. In this case, $k$ is an integer and cannot be deformed
continuously, hence the bulk dual contains
no dilaton.  For the non-supersymmetric theory, it is not immediately obvious that the ``M-theory'' large-$N$
limit  exists; however, if the limit does
exist, the holographic dual would not contain a dilaton or axion, and may be as
simple as Einstein-Maxwell gravity with a small negative cosmological constant. 

We might instead consider theories based on gauge group $O(N) \times O(N)$,
coupled to `bi-vector' \textit{Majorana} fermions instead of
bifundamental Dirac fermions. In this theory, the spin-1 current $\bar{\psi}
\gamma^\mu \psi$ vanishes identically because $\psi$
is a Majorana fermion\footnote{Recall that $\bar{\chi}\gamma^\mu \psi = - \bar{\psi}\gamma^\mu \chi$ for Majorana fermions $\chi$ and $\psi$}, as do all other odd-spin conserved currents $J_s$ in the
free theory. The only conserved and protected current operator
appears to be the spin-2 current . Hence 
if the finite-$k$, large-$N$ limit exists for the $O(N) \times O(N)$ theory, the holographic dual may be as simple as \textit{pure} Einstein gravity with a small negative cosmological constant. (While the standard 't Hooft limit may also include a dilaton.) 

Bifundamental theories are large-$N$ ``matrix models'', which are as intractable at strong coupling  as classic large-$N$
theories with adjoint degrees of freedom. Without supersymmetry, studying these
theories at strong coupling appears impossible; making tests of any conjectural
holographic duals difficult. However, an indirect way of
obtaining information about the large $N$ limit of $U(N) \times U(N)$ theories at strong coupling was proposed in \cite{Chang:2012kt}. The crucial idea is to instead consider a $U(N) \times U(M)$ theory (similar to \cite{Aharony:2008gk}), with $M \ll N$, in an $M/N$ expansion. $U(N)$ (or $U(N)|_k \times U(1)|_{-k}$) Chern-Simons theory coupled to fundamental matter is effectively a vector model, and hence exactly (at least in  principle)
solvable in the 't Hooft limit \cite{Giombi:2011kc, GurAri:2012is,
Aharony:2012nh}. Studying the $U(N) \times U(M)$ theory in an
$M/N$ expansion allows us to perturbatively approach a strongly interacting
matrix model starting from strongly interacting but exactly solvable vector
model. 

In this paper, we consider $U(N)
\times U(M)$ Chern-Simons coupled to bifundamental fermions, with levels
$k_N$ and $k_M$ respectively. We work in the large $N$ limit, keeping the
following three parameters fixed: $\lambda_N = \frac{N}{k_N}$, $\lambda_M =\frac{N}{k_M}$, and $\alpha = \frac{M}{N}$. We always assume $M\leq N$.
%

Important data for any conjecture about holographic duals are the scaling
dimensions $\Delta$ of single-trace primary operators, which are directly related to the masses of the fields in the dual bulk description. For operators of spin $s>0$, unitarity requires that $\Delta \geq s+1$ \cite{Mack:1975je, Minwalla:1997ka}. Massless fields of spin $s>0$ in the bulk are dual to primary operators with $\Delta=s+1$, which are in short representations of the conformal group and hence are conserved currents. It is easy to convince oneself, following \cite{Giombi:2011kc, Gaiotto:2007qi}, that the free theory contains an infinite tower of conserved currents, one for each spin $s \geq 1$, of the schematic form $\sim \tr \bar{\psi} \gamma_{\mu} \partial^{s-1}\psi$. When $M=1$ and $N$ is large, the scaling dimensions of these operators (which in addition to the scalar $\bar{\psi}\psi$, which is in a long representation, are the only single trace primary operators in the $M=1$ theory) are protected by a simple argument based on conformal representation theory, given in \cite{Giombi:2011kc, Aharony:2011jz} -- the holographic dual is therefore believed to be a Vasiliev theory (see \cite{Vasiliev:1999ba, Klebanov:2002ja, Sezgin:2002rt, Chang:2012kt, Maldacena:2011jn, Maldacena:2012sf} and \cite{Giombi:2012ms} for a review) with an infinite tower of higher-spin gauge fields even at strong coupling. However, when $M=N$ these operators are now part of a much larger family of single-trace primary operators. With the exception of the spin-1 and spin-2 currents, these operators are unprotected and do acquire anomalous dimensions as one can verify explicitly by a one-loop
calculation of, say, the spin-3 current. 

These anomalous dimensions, as well as those of other generic unprotected operators, would have to diverge at strong coupling if the theory has a traditional holographic dual based on Einstein-gravity. It would be particularly interesting if we could obtain any information about the anomalous dimension of any unprotected operator in some strong coupling limit.

As we will show in this paper, it is possible to calculate anomalous dimensions of generic operators to all orders in $\lambda$ in an $M/N$ expansion. The simplest unprotected single-trace primary operator is the scalar $\tr \bar{\psi}\psi$. As a preliminary study of the theory, we will calculate its anomalous dimension in what appears to be the simplest, potentially non-trivial limit: We set $\lambda_N = 0$, and work to
first order in $\alpha=M/N$, but to all orders in $\lambda_M$. This is similar to a large flavor expansion, common in condensed matter physics.

Though first-order results in an $M/N$ expansion are similar to $1/N$ calculations in a vector model, at higher orders, the $M/N$ expansion is different from and substantially simpler than a $1/N$ expansion. $M/N=1$ simply corresponds to a large $N$ saddle point of a $U(N) \times U(N)$ (or $O(N)\times O(N)$) theory -- so at any stage we only have to include \textit{planar} Feynman diagrams -- by planar we mean those that can be drawn on a genus-zero surface \cite{'tHooft:1973jz} -- in our calculations. When $M/N$ is very small, we need to draw only the relatively small (but infinite) subset of planar diagrams that contribute to the leading large-$N$ solution of a vector model -- as shown in, e.g. \cite{Giombi:2011kc}, one can explicitly perform a sum over these diagrams. The size of this planar subset increases as we consider higher order corrections in $M/N$; but it is still possible to sum over all the diagrams by solving a finite number of integral equations. When $M/N$ is order unity, we need to include all planar diagrams, because all planar diagrams contribute to the leading large-$N$ solution of a ``matrix'' model containing adjoint or bifundamental degrees of freedom -- performing a sum over all planar diagrams appears impossible. 


\section{The Anomalous Dimension of $\bar{\psi}\psi$}
We work in the Euclidean theory, where the Chern-Simons action takes the form:
\begin{eqnarray*}
 S^{\text{euc}}_{\text{CS}} & = & \frac{i}{4\pi} \int
d^3x~\epsilon^{\mu\nu\lambda} \left[k_M \text{tr}\left(A_\mu \partial_\nu A_\lambda
+ \frac{2}{3} A_\mu A_\nu A_\lambda \right) - k_N \text{tr}\left(B_\mu \partial_\nu
B_\lambda + \frac{2}{3} B_\mu B_\nu B_\lambda \right) \right] 
\end{eqnarray*}
  We work in light-cone gauge, $A_{-}=B_{-}=0$; in this gauge, ghosts decouple and cubic interactions are not present. (Let $x^1$, $x^2$ and $x^3$ be coordinates in Euclidean signature, we define light-cone coordinates as $x^+\equiv (x^1+ix^2)/\sqrt{2}=x_-$ and $x^-\equiv (x^1-ix^2)/\sqrt{2}=x_+$. We also define $x_s^2=x_1^2+x_2^2=2x_+x_-$. )

The fermion action is 
\begin{equation}
 \int d^3x \tr \bar{\psi}\gamma^\mu D_\mu  \psi,
\end{equation}
and the $\gamma^i$ can be taken to be the Pauli matrices $\sigma^i$. Covariant derivatives are 
$$D_\mu \psi = \partial_\mu \psi -i A_\mu \psi + i \psi B_\mu,$$
as usual for a bifundamental theory. 

The gauge propagator $D_{\mu\nu}$ is nonzero only for $D_{+3}(p)=-D_{3+}= \frac{4\pi}{ik} \frac{1}{p^+}$ and the fermion propagator is $\frac{1}{i{p_\mu \gamma^\mu}}$. (Here we suppress group  theory factors. Our normalization conventions for gauge group generators is $\tr T^aT^b=\frac{1}{2}\delta^{ab}$.)

Since we set $\lambda_N=0$, we will drop the subscript $M$ on $\lambda_M$. In this limit, which is quite different from the limit considered in \cite{Giombi:2011kc}, our theory is essentially a $U(M)$ theory with $N$ flavors (but in the singlet sector, as the $U(N)$ symmetry group is still gauged), which is trivial at zeroth order. The calculation of the anomalous dimension in this limit turns out to be very
similar to some of the two-loop $1/N$ calculations in \cite{Giombi:2011kc}. See also \cite{Gaiotto:2007qi} for similar two-loop calculations in the case of superconformal Chern-Simons theories.

\subsection{Scalar Two-point function}
 The two point function of the scalars, which has scaling dimension $\Delta=2$, in free theory is given by:
 \begin{eqnarray}
-NM \tr \int \frac{d^3p}{(2\pi)^3} \frac{1}{i p_\mu \gamma^\mu} \frac{1}{i
 (p_\nu+q_\nu) \gamma^\nu} = NM\frac{1}{2\pi} \Lambda -NM\frac{1}{8}|q|
 \end{eqnarray} 
We calculate the anomalous dimension $\delta=\Delta-2$ by extracting the coefficient of logarithmic divergences of the two-point function in the interacting theory. Our regularization procedure is that used in, e.g., \cite{Giombi:2011kc, GurAri:2012is, Aharony:2012ns} -- we place a hard momentum cutoff $\Lambda$ on momentum integrals in the 1-2 plane.

\subsection{Gauge Propagator}
To first order in $M/N$ the gauge propagator receives an infinite series of self-energy corrections. These are given by:

\begin{fmffile}{gaugex}
        \begin{tabular}{c}
            \begin{fmfgraph*}(60,40)
                \fmfleft{i1}
                \fmfright{o1}
 \fmfv{decoration.shape=circle, decoration.size=.2h, decoration.filled=shaded}{z}
		 \fmf{photon,tension=.5}{i1,z,o1}
             \end{fmfgraph*} 
        \end{tabular}
 \end{fmffile}
	$=$ 
\begin{fmffile}{gauge0}
        \begin{tabular}{c}
            \begin{fmfgraph*}(40,40)
                \fmfleft{i1}
                \fmfright{o1}
 		 \fmf{photon,tension=5}{i1,o1}
             \end{fmfgraph*}
        \end{tabular}
        \end{fmffile}
$+$
\begin{fmffile}{gauge1}
        \begin{tabular}{c}
            \begin{fmfgraph*}(60,40)
                \fmfleft{i1}
                \fmfright{o1}
		\fmf{fermion,left, tension=2}{a,b,a}
		\fmf{photon,tension=5}{i1,a}
		\fmf{photon, tension=5}{b,o1}
             \end{fmfgraph*}
        \end{tabular}
        \end{fmffile}
$+$
\begin{fmffile}{gauge2}
        \begin{tabular}{c}
            \begin{fmfgraph*}(80,40)
                \fmfleft{i1}
                \fmfright{o1}
		\fmf{fermion,left, tension=1.5}{a,b,a}
	        \fmf{fermion,left, tension=1.5}{c,d,c}
		\fmf{photon,tension=5}{i1,a}
		\fmf{photon, tension=5}{b,c}
		\fmf{photon,tension=5}{d,o1}
             \end{fmfgraph*}
\end{tabular}
$ + ~ \ldots $.
\end{fmffile}

The gauge field's one-loop self energy, $\Sigma^{\mu\nu}$ is given by 
\begin{equation}
 \Sigma^{\mu\nu}(q)  = -\frac{N}{2} (-1)\tr \int \frac{d^3p}{(2\pi)^3}
\frac{1}{i p_\alpha \gamma^\alpha} \gamma^\mu \frac{1}{i
(p_\beta+q_\beta) \gamma^\beta} \gamma^\nu = -N \frac{q^2 \delta^{\mu\nu} - q^\mu
q^\nu}{32 q}
\end{equation}

Let $D_{\mu\nu}$ be the free gauge propagator. Let $G_{\mu\nu}$ be the gauge
propagator with an infinite series of self-energy corrections. (Both are
$3\times 3$ matrices -- though they are effectively
$2\times2$ matrices in light cone gauge). Then, the matrices $\mathbf{G}$ and $\mathbf{D}$ satisfy 
$
 \mathbf{G} = \left( 1 - \mathbf{D}\mathbf{\Sigma} \right)^{-1} \mathbf{D}
$.

Evaluating this we have
\begin{equation}
\mathbf{G} = \begin{pmatrix} G_{33} & G_{3+} \\ G_{+3} & G_{++}
              \end{pmatrix}
= \frac{1}{N} \frac{2\pi^2q_+^2}{q q_s^4} \frac{64}{64+\pi^2 \lambda^2}
\begin{pmatrix} \lambda^2 q_-^2  & \frac{8i\lambda}{\pi} q_- q - \lambda^2 q_-
q_3 \\
 -\frac{8i\lambda}{\pi} q_- q - \lambda^2 q_- q_3  & -q_s^2 \lambda^2
 \end{pmatrix}
\end{equation}

\subsection{Calculation}
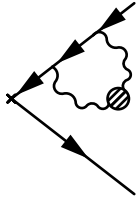
\begin{figure}
\center
\begin{fmffile}{Vertex1}
        \begin{tabular}{c}
            \begin{fmfgraph*}(60,80)
                \fmfleft{i1}
                \fmfright{o1,o2}
                \fmffixed{(-.1h,0)}{y1,i1}
                \fmffixed{(0,-.05h)}{y2,o1}
                \fmffixed{(0,.05h)}{y3,o2}
                \fmfv{decoration.shape=cross, decoration.size=.05h,
decoration.filled=empty}{y1}
                \fmffixed{(0,.5h)}{o1,v}
                \fmffixed{(0,.5h)}{v,o2}
                \fmffixed{(-.5h,0)}{z,y1}
                \fmf{photon, left=.6,tension=.0}{x,z,y}
                \fmfv{decoration.shape=circle, decoration.size=.1h,
decoration.filled=shaded}{z}
                \fmf{electron}{y3,x,y,y1,y2}
             \end{fmfgraph*}
        \end{tabular}
        \end{fmffile} 
\caption{
Correction to the scalar vertex due to the $O(M/N)$ fermion self-energy}
\label{self}
\end{figure}

\begin{figure}
\center
 \begin{fmffile}{Vertex2}
        \begin{tabular}{c}
            \begin{fmfgraph*}(60,80)
                \fmfleft{i1}
                \fmfright{o1,o2}
                \fmffixed{(-.1h,0)}{y1,i1}
                \fmffixed{(0,-.05h)}{y2,o1}
                \fmffixed{(0,.05h)}{y3,o2}
               \fmfv{decoration.shape=cross, decoration.size=.05h,
decoration.filled=empty}{y1}
                \fmffixed{(0,.5h)}{o1,v}
                \fmffixed{(-.45h,0)}{z,y1}
                \fmffixed{(0,.5h)}{v,o2}
                \fmf{photon,right=.5,tension=.01}{x,z,y}
                \fmfv{decoration.shape=circle, decoration.size=.1h,
decoration.filled=shaded}{z}
                \fmf{electron}{y3,y,y1,x,y2}
             \end{fmfgraph*}
        \end{tabular}
        \end{fmffile}
\caption{Another correction to the vertex.}
\label{diag2}
\end{figure}
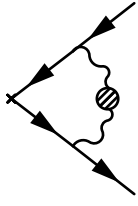

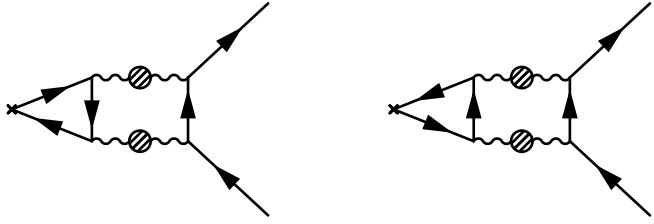
\begin{figure}
\center
\begin{tabular}{cc}
\begin{fmffile}{Additional}
            \begin{fmfgraph*}(120,80)
                \fmfleft{ii1}
		\fmfright{oo1,oo2}
		\fmffixed{(.05w,0)}{ii1,i1}
		\fmffixed{(.05w,0)}{o1,oo1}
		\fmffixed{(.05w,0)}{o2,oo2}
		\fmffixed{(0,.5h)}{o1,v}
		\fmffixed{(0,.5h)}{v,o2}	
		\fmffixed{(.3w,0)}{x,w}
		\fmffixed{(.3w,0)}{y,z}
		\fmffixed{(0,-.3h)}{x,y}
		\fmf{photon,tension=1}{x,a,w}
		 \fmfv{decoration.shape=circle, decoration.size=.1h, decoration.filled=shaded}{a}
		  \fmfv{decoration.shape=circle, decoration.size=.1h, decoration.filled=shaded}{b}
		\fmf{photon,tension=1}{y,b,z}
                \fmf{electron,  tension=.01}{i1,x,y,i1}
                \fmf{electron,  tension=.01}{o1,z,w,o2}
		 \fmfv{decoration.shape=cross, decoration.size=.05h, decoration.filled=empty}{i1}
\end{fmfgraph*}
\end{fmffile}
& 
\begin{fmffile}{Additional2}
\begin{fmfgraph*}(120,80)
        \fmfleft{ii1}
        \fmfright{oo1,oo2}
        \fmffixed{(.05w,0)}{ii1,i1}
        \fmffixed{(.05w,0)}{o1,oo1}
        \fmffixed{(.05w,0)}{o2,oo2}
        \fmffixed{(0,.5h)}{o1,v}
        \fmffixed{(0,.5h)}{v,o2}    
        \fmffixed{(.3w,0)}{x,w}
        \fmffixed{(.3w,0)}{y,z}
        \fmffixed{(0,-.3h)}{x,y}
        \fmf{photon,tension=1}{x,a,w}
         \fmfv{decoration.shape=circle, decoration.size=.1h, decoration.filled=shaded}{a}
          \fmfv{decoration.shape=circle, decoration.size=.1h, decoration.filled=shaded}{b}
        \fmf{photon,tension=1}{y,b,z}
                \fmf{electron,  tension=.01}{i1,y,x,i1}
                \fmf{electron,  tension=.01}{o1,z,w,o2}
         \fmfv{decoration.shape=cross, decoration.size=.05h, decoration.filled=empty}{i1}
\end{fmfgraph*}
\end{fmffile}
\end{tabular}
\caption{
Additional contribution to the two-point function.}
\label{additional}
\end{figure}

The diagrams which contribute to the anomalous dimension in our
limit are shown in Figure \ref{self}, \ref{diag2}, and \ref{additional}. 

The corrected self energy of the fermion, $\Sigma_\psi(p)$, needed to evaluate the diagram in
Figure \ref{self} is given by
\begin{equation}
  \Sigma_\psi(p) = {M\over 2} \int \frac{d^3q}{(2\pi)^3}
\frac{-i\gamma^\mu({\slash\!\!\!
p}+{\slash\!\!\! q})\gamma^\nu}{(p+q)^2} G_{\mu\nu}(q)
\end{equation}

Evaluating this diagram we have,
\begin{equation}
\gamma^\mu {\slash\!\!\! p} \gamma^\nu G_{\mu\nu} = (G_{+3} +
G_{3+})(p_3 \gamma^+ + p_- \gamma^3) + (G_{+3} - G_{3+}) p_- + 2 G_{++} p_-
\gamma^+ + G_{33} (p_3 \gamma^3 - p_- \gamma^- - p_+ \gamma^+)
\end{equation}
hence,
\begin{equation}
  \Sigma_\psi(p) = -i{M\over N } \frac{64\pi^2}{64+\pi^2\lambda^2} \int {d^3q\over
(2\pi)^3} \frac{1}{(p+q)^2} (K_\mu \gamma^\mu + K_I)
\end{equation}
where
\begin{eqnarray}
 K_- & = & -\frac{(p_- + q_-)  \lambda^2}{4 q} \\
 K_+ & =  & -(p_3+q_3) \lambda^2 \frac{q_+ q_3}{q q_s^2} + (p_+ + q_+) \frac{1}{4
q}
\lambda^2 - 2 p_-  \frac{q_+^2}{q q_s^2} \lambda^2 -   \frac{q_+}{q }
\lambda^2\\
 K_3 & = & \frac{(p_3+q_3)  \lambda^2}{4 q} -  p_-  \lambda^2 \frac{q_3q_+} {q
q_s^2} - \lambda^2 \frac{q_3} {2 q } \\
 K_I & = &  - p_- \frac{8 i \lambda }{\pi}\frac{q_+}{q_s^2} - \frac{4 i \lambda
}{\pi}
\end{eqnarray}

To evaluate these integrals, we use Feynman parameters:
\begin{eqnarray*}
\int \frac{d^3q}{(2\pi)^3} \frac{f(q)}{q (p+q)^2} & = & \frac{1}{2} \int_0^1 dx
\int \frac{d^3q}{(2\pi)^3} (1-x)^{-1/2} \frac{f(q-xp)}{(q^2+x(1-x)p^2)^{3/2}} 
\\
 \int \frac{d^3q}{(2\pi)^3} \frac{f(q_3, \vec{q_s})}{q_s^2 (p+q)^2} & = &
\int_0^1 dx \int \frac{d^3q}{(2\pi)^3} \frac{f(q_3-p_3, \vec{q}_s - x
\vec{p}_s)}{(q_s^2+x(1-x)p_s^2 + q_3^2)^2} 
\\
 \int \frac{d^3q}{(2\pi)^3} \frac{f(q_3, \vec{q_s})}{q_s^2 q (p+q)^2} & = &
\frac{3}{4}\int_0^1 dy \int_0^{1-y}dz \int \frac{d^3q}{(2\pi)^3} 
 y^{-1/2}\frac{f(q_3-\frac{z}{y+z}p_3, \vec{q}_s - z \vec{p}_s)}
{\left(q_s^2+z(1-z)p_s^2 + (y+z)q_3^2 + \frac{yz}{y+z}p_3^2\right)^{5/2}}
\end{eqnarray*}

Keeping only the logarithmic divergences even in $\lambda$, the
integrals we need to evaluate are:
 \begin{eqnarray*}
&& -\frac{\lambda^2}{2} \int_0^1 dx \int \frac{d^3q}{(2\pi)^3} (1-x)^{-1/2}
\frac{(1-x) p_\mu \gamma^\mu - 2 p_3 \gamma^3 -4 x p_+ \gamma^+}{4
(q^2+x(1-x)p^2)^{3/2}} \\
& = & -\frac{\lambda^2}{6 (2\pi)^2} \left( p_\mu \gamma^\mu - 6 p_3 \gamma^3 - 8
p_+ \gamma^+ \right) \ln \Lambda
\end{eqnarray*}
and
\begin{eqnarray*}
&& -\frac{3\lambda^2}{4} \int_0^1 dy \int_0^{1-y}dz \int \frac{d^3q}{(2\pi)^3} 
y^{-1/2} \frac{-z q_3^2 p_+ \gamma^+ } {\left(q_s^2+z(1-z)p_s^2 + (y+z)q_3^2 +
\frac{yz}{y+z}p_3^2\right)^{5/2}} \\
& = & \frac{\lambda^2}{6\pi^2} p_+ \gamma^+ \ln \Lambda
\end{eqnarray*}

Putting these together, we have  
\begin{equation}
\Sigma_\psi(p) = -i{M\over N }
\frac{64\pi^2}{64+\pi^2\lambda^2}\frac{\lambda^2}{6 (2\pi)^2} \left( -p_\mu
\gamma^\mu + 6 p_3 \gamma^3 + 12 p_+ \gamma^+ \right) \ln \Lambda + {\rm const}
\end{equation}
Including this in a two-point function calculation gives a contribution to the
anomalous dimension $\delta_1 =-\frac{M}{N} \frac{40
\lambda^2}{3(64+\pi^2\lambda^2)}$.

The other diagram that contributes is Figure \ref{diag2}, and is given by 
\begin{equation}
 -\frac{M}{2}\int \frac{d^3q}{(2\pi)^3} \frac{1}{q^2}G_\mu{}^\mu(q) = \delta_2 \ln
\Lambda + \text{ const}
\end{equation}

We have:
\begin{equation}
 G_{\mu}{}^{\mu} = \frac{1}{Nq} \frac{32 \pi^2 \lambda^2}{\pi^2\lambda^2+64}
\end{equation}
and the contribution to the anomalous dimension is 
\begin{equation}
  \delta_2=-\frac{M}{N} \frac{8\lambda^2}{\pi^2 \lambda^2 + 64}
\end{equation}

Next, we evaluate the logarithmic divergences arising from the additional diagrams shown in Figure \ref{additional}.
The sum of the two\footnote{We thank Aaron Hui for helping to correct an earlier version of this paper.} diagrams is given by the following 
\begin{equation*}
    I_3 =\frac{1}{2}\tr \left( \left(-2 \frac{MN}{4}\right)\int \frac{d^3 q}{(2\pi)^3} \gamma^\mu \frac{1}{i(\slashed{p}-\slashed{q})} \gamma^\nu G_{\mu\alpha}(q) G_{\beta \nu}(q) C^{\alpha \beta}(q)\right)
\end{equation*}
where
\begin{equation*}
    C^{\mu \nu}(q) = \int \frac{d^3p}{(2\pi)^3} \Tr \left( \frac{1}{i (\slashed{p}-\slashed{q})} \gamma^\mu \frac{1}{i \slashed{p}}  \frac{1}{i \slashed{p}} \gamma^\nu \right) 
\end{equation*}

We find $I_3 = \delta_3 (-\log \Lambda)$
with
\begin{equation}
\delta_3=-\frac{64\lambda^2}{64+\pi^2\lambda^2}  \frac{ 64 - \lambda^2\pi^2}{ 64 + \lambda^2 \pi^2} \frac{M}{N}
\end{equation}

\begin{figure}
\center
 \includegraphics[scale=.5]{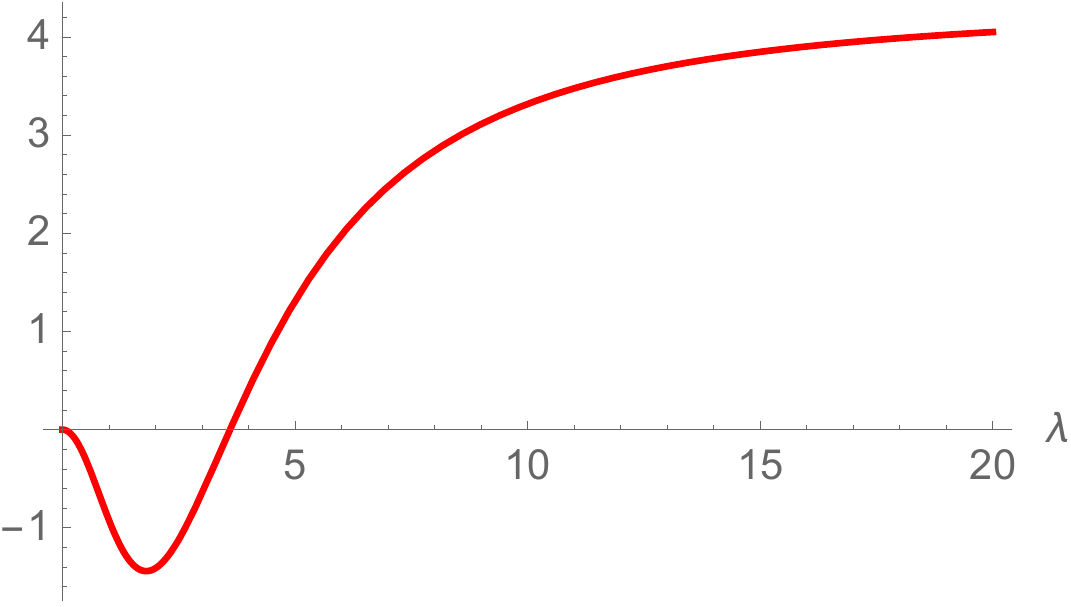} 
  \caption{The anomalous dimension $\delta/(\frac{M}{N})$ of $\tr \bar{\psi}\psi$, from equation \eqref{result} as a function of $\lambda$. It reaches a minimum of $\delta=-\frac{128}{9\pi^2} \frac{M}{N}$ at $\lambda \pi= 4 \sqrt{2}$. The limiting value as $\lambda \rightarrow \infty$ is $\delta \rightarrow \frac{128}{3\pi^2} \frac{M}{N}$. \label{plota}}
\end{figure}

The anomalous dimension is then given by $\delta=\delta_1+\delta_2+\delta_3$, which is
\begin{equation}
\delta=\frac{128 \lambda ^2 \left(\pi
   ^2 \lambda ^2-128\right)}{3
   \left(\pi ^2 \lambda
   ^2+64\right)^2} \frac{M}{N}. \label{result}
\end{equation} 
This is plotted in Figure \ref{plota}.

We can also carry out a perturbative calculation in Feynman gauge as in \cite{Giombi:2011kc}; working to order $\lambda^2$ we find agreement with the above result.

\section{Discussion}

\subsection{Strong-Coupling Limit}
We find the anomalous dimension, though unprotected,
does not diverge at strong coupling and instead approaches a finite value,
\begin{equation}
 \delta \rightarrow \frac{M}{N}\frac{128}{3\pi^2}, \label{sc}
\end{equation}
as $\lambda_M \rightarrow \infty$. Based on the form of the exact gauge propagator; we expect that at any given order in $M/N$ the anomalous dimension would be finite in the 't Hooft limit for all values of $\lambda$. Only as $M/N \rightarrow 1$, can we expect the anomalous dimension may diverge for some finite value of $\lambda$ or as $\lambda \rightarrow \infty$. 



As $\lambda \rightarrow \infty$ the theory becomes parity-preserving. Hence one might expect that it should be equivalent to say, the IR limit of QCD in three dimensions with gauge group $SU(M)$ and $N_f$ flavours. This conjecture is perhaps analogous to statements about the IR limits of super-Yang-Mills theories in three dimensions and supersymmetric bifundamental theories such as ABJM\cite{Schwarz:2004yj, Bagger:2007jr, VanRaamsdonk:2008ft, Aharony:2008ug}. Calculations of the anomalous dimension in the IR limit of QCD${}_3$ would be similar to calculations in QED${}_3$  carried out in, e.g., \cite{PhysRevB.66.144501, Pufu:2013vpa} (see also \cite{Appelquist:1988sr}) (but, of course, ghosts and cubic interaction vertices have to be considered) and it would be interesting to investigate this further.


\subsection{All orders in $\lambda_N$}
Studying the theory at $\lambda_N=0$ drastically simplified our calculations. It is possible in principle to calculate the $M/N$ corrections to the anomalous
dimension to all orders in $\lambda_N$ as well. 

Let us show diagrammatically how one can calculate the anomalous dimension to all orders in $\lambda_N$, with $\lambda_M=0$ for simplicity, so we only need draw one gauge field. (For an all orders in both $\lambda_N$ and $\lambda_M$, we would also need to take into account the various $\lambda_N$ corrections in the diagrams shown in Figures 1, 2 and 3). We will need the exact fermion propagator:
\begin{equation}
\begin{fmffile}{propx}
        \begin{tabular}{c}
            \begin{fmfgraph*}(60,40)
                \fmfleft{i1}
                \fmfright{o1}
 \fmfv{decoration.shape=square, decoration.size=.2h, decoration.filled=shaded}{z}
		 \fmf{plain,tension=.5}{i1,z,o1}
             \end{fmfgraph*} 
        \end{tabular}
 \end{fmffile}
	= 
\begin{fmffile}{prop0}
        \begin{tabular}{c}
            \begin{fmfgraph*}(60,40)
                \fmfleft{i1}
                \fmfright{o1}
 		 \fmf{electron,tension=5}{i1,o1}
             \end{fmfgraph*}
        \end{tabular}
        \end{fmffile}
+
\begin{fmffile}{prop1}
        \begin{tabular}{c}
            \begin{fmfgraph*}(120,40)
                \fmfleft{i1}
                \fmfright{o1}
		\fmf{photon,left, tension=2}{a,b}
		\fmf{electron,tension=5}{i1,a}
                \fmf{plain,tension=5}{a,z,b}
                \fmf{plain,tension=5}{b,c,o1}
\fmfv{decoration.shape=square, decoration.size=.13h, decoration.filled=shaded}{z}

\fmfv{decoration.shape=square, decoration.size=.13h, decoration.filled=shaded}{c}

             \end{fmfgraph*}
        \end{tabular}
        \end{fmffile},
\end{equation}
the exact quantum corrected vertex:
\begin{equation}
\begin{fmffile}{ex_vert_1}
        \begin{tabular}{c}
            \begin{fmfgraph*}(60,40)
                \fmfleft{v}
                \fmfright{o1,o2}
                \fmfv{decoration.shape=circle, decoration.size=.2h, decoration.filled=30}{v}
                \fmf{plain_arrow}{o2,v,o1}
             \end{fmfgraph*}
        \end{tabular}
        \end{fmffile}
=
\begin{fmffile}{ex_vert_0}
        \begin{tabular}{c}
            \begin{fmfgraph*}(60,40)
                \fmfleft{v}
                \fmfright{o1,o2}
                \fmfv{decoration.shape=cross, decoration.size=.2h}{v}
                \fmf{plain_arrow}{o2,v,o1}
             \end{fmfgraph*}
        \end{tabular}
        \end{fmffile}
+
\begin{fmffile}{ex_vert_2}
        \begin{tabular}{c}
            \begin{fmfgraph*}(60,40)
                \fmfleft{v}
                \fmfright{o1,o2}
                \fmfv{decoration.shape=circle, decoration.size=.2h, decoration.filled=30}{v}
                \fmf{plain}{o2,x,v,y,o1}
\fmfv{decoration.shape=square, decoration.size=.13h, decoration.filled=shaded}{x}
\fmfv{decoration.shape=square, decoration.size=.13h, decoration.filled=shaded}{y}
		\fmf{photon}{o2,o1}
             \end{fmfgraph*}
        \end{tabular}
        \end{fmffile},
\end{equation}
 and the ``ladder'' diagrams\footnote{We thank A. Bedhotiya for discussions on this and attempts to evaluate the ladder diagram.}:
\begin{equation}
\begin{fmffile}{ladder}
\begin{tabular}{c}
   \begin{fmfgraph*}(120,50)
                \fmfleft{i1,i2}
                \fmfright{o1,o2}
		\fmf{electron}{i1,x,a,o1}
		\fmf{electron}{o2,y,b,i2}
		\fmfpoly{filled=hatched}{x,a,y,b}
             \end{fmfgraph*}
 \end{tabular}
\end{fmffile}
=
\begin{fmffile}{ladder0}\begin{tabular}{c}
   \begin{fmfgraph*}(90,50)
                \fmfleft{i1,i2}
                \fmfright{o1,o2}
		\fmf{electron}{i1,x,o1}
		\fmf{electron}{o2,y,i2}
		\fmf{photon}{x,y}
             \end{fmfgraph*}
 \end{tabular}\end{fmffile}
+
\begin{fmffile}{ladder1}\begin{tabular}{c}
   \begin{fmfgraph*}(160,65)
                \fmfleft{i1,i2}
                \fmfright{o1,o2}
		\fmf{electron}{i1,x,a} 
		\fmf{plain, tension=2}{a,w1,c} 
		\fmf{electron}{c,o1}
		\fmf{electron}{o2,d} 
		\fmf{plain, tension=2}{d,w2,y}
		\fmf{electron}{y,b,i2}
		\fmfpoly{filled=hatched}{x,a,y,b}
		\fmfv{decoration.shape=square, decoration.size=8, decoration.filled=shaded}{w1}
	  \fmfv{decoration.shape=square, decoration.size=8, decoration.filled=shaded}{w2}
	        \fmf{photon}{c,d}
             \end{fmfgraph*}
 \end{tabular}
\end{fmffile}
\end{equation}

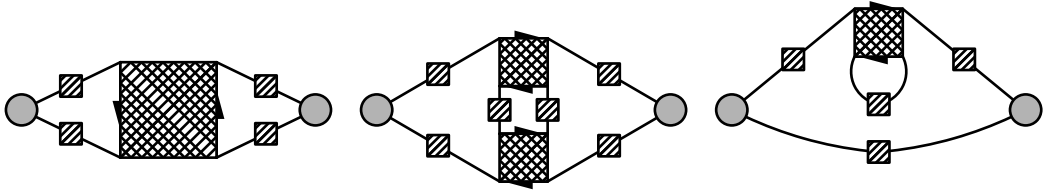
\begin{figure}
\begin{tabular}{ccc}
\begin{fmffile}{anomb}
 \begin{fmfgraph*}(110,90)
\begin{tabular}{c}
 \fmfleft{i} 
 \fmfright{o}
\fmffixed{(0,-.4h)}{b,c}
\fmfpoly{filled=hatched}{a,b,c,d}
  \fmf{electron}{b,c} 
  \fmf{electron}{d,a}
  \fmf{plain}{i,w1,b,c,w2,i} 
  \fmf{plain}{a,w3,o,w4,d,a}
 \fmfv{decoration.shape=circle, decoration.size=.13h, decoration.filled=30}{i}
\fmfv{decoration.shape=circle, decoration.size=.13h, decoration.filled=30}{o}  
\fmfv{decoration.shape=square, decoration.size=8, decoration.filled=shaded}{w1}
\fmfv{decoration.shape=square, decoration.size=8, decoration.filled=shaded}{w2}
\fmfv{decoration.shape=square, decoration.size=8, decoration.filled=shaded}{w3}
\fmfv{decoration.shape=square, decoration.size=8, decoration.filled=shaded}{w4}
\end{tabular}
\end{fmfgraph*} 
\end{fmffile}
&
\begin{fmffile}{anomb2}
 \begin{fmfgraph*}(110,90)
\begin{tabular}{c}
 \fmfleft{i} 
 \fmfright{o}
\fmffixed{(0,-.2h)}{b,c}
\fmffixed{(0,-.2h)}{c,b2}
\fmffixed{(0,-.2h)}{b2,c2}
\fmfpoly{filled=hatched,tension=2}{a,b,c,d}
\fmf{electron}{b,a}
\fmf{electron}{d,c}
\fmfpoly{filled=hatched,tension=2}{a2,b2,c2,d2}
\fmf{electron}{b2,a2}
\fmf{electron}{d2,c2}
\fmf{plain,tension=.5}{a2,w1,d,c,w2,b2,a2}
  \fmf{plain}{i,w3,b,a,w4,o,w5,d2,c2,w6,i}
 \fmfv{decoration.shape=circle, decoration.size=.13h, decoration.filled=30}{i}
\fmfv{decoration.shape=circle, decoration.size=.13h, decoration.filled=30}{o}  
\fmfv{decoration.shape=square, decoration.size=8, decoration.filled=shaded}{w1}
\fmfv{decoration.shape=square, decoration.size=8, decoration.filled=shaded}{w2}
\fmfv{decoration.shape=square, decoration.size=8, decoration.filled=shaded}{w3}
\fmfv{decoration.shape=square, decoration.size=8, decoration.filled=shaded}{w4}
\fmfv{decoration.shape=square, decoration.size=8, decoration.filled=shaded}{w5}
\fmfv{decoration.shape=square, decoration.size=8, decoration.filled=shaded}{w6}
\end{tabular}
\end{fmfgraph*} 
\end{fmffile}
&
\begin{fmffile}{anomb3}
 \begin{fmfgraph*}(110,90)
\begin{tabular}{c}
 \fmfleft{i} 
 \fmfright{o}
\fmffixed{(0,-.2h)}{b,c}
\fmffixed{(0,-.2h)}{c,b2}
\fmffixed{(0,-.2h)}{b2,c2}
\fmffixed{(-0.1h,0)}{d2,d3}
\fmffixed{(0,-.2h)}{w2,d3}
\fmfpoly{filled=hatched,tension=2}{a,b,c,d}
\fmfpoly{phantom,tension=2}{a2,b2,c2,d2}
\fmf{electron}{d,c}
\fmf{plain,right=.5,tension=.1}{c,w2,d}
\fmf{electron}{b,a}
\fmf{plain}{i,w3,b,a,w4,o}
  
  \fmf{plain,left=.1,tension=1.2}{o,d3,i}
 \fmfv{decoration.shape=circle, decoration.size=.13h, decoration.filled=30}{i}
\fmfv{decoration.shape=circle, decoration.size=.13h, decoration.filled=30}{o}  

\fmfv{decoration.shape=square, decoration.size=8, decoration.filled=shaded}{d3}
\fmfv{decoration.shape=square, decoration.size=8, decoration.filled=shaded}{w2}
\fmfv{decoration.shape=square, decoration.size=8, decoration.filled=shaded}{w3}
\fmfv{decoration.shape=square, decoration.size=8, decoration.filled=shaded}{w4}

\end{tabular}
\end{fmfgraph*} 
\end{fmffile}
\end{tabular}
\caption{
The two point function of the scalar primary to all orders in $\lambda_N$.} \label{ln}
\end{figure}

In terms of these ingredients -- the first two of which are calculated in \cite{Giombi:2011kc, GurAri:2012is}, and the third is extensively discussed in \cite{Jain:2014nza} which recently appeared -- the diagrams that contribute to the two-point function are shown in in Figure \ref{ln}. Evaluating the diagrams in Figure \ref{ln} is more challenging from a technical point of view, primarily due to the exact planar off-shell fermion four-point function.  While we hope to carry out such a calculation in the near future, we suspect that the anomalous dimensions of unprotected operators would remain finite at strong coupling unless $\alpha=M/N=1$ or some other finite value. 

\subsection{Higher-Spin Operators and Other Calculations}

We chose to work with the scalar $\tr \bar{\psi}\psi$ as the simplest representative of an unprotected operator -- however, even if the theory possesses a simple holographic dual, the scalar's anomalous dimension need not diverge if $M/N \rightarrow 1$; we only require higher spin operators to acquire large anomalous dimensions.   It would be interesting to calculate the first-order $M/N$ anomalous dimension for one of the higher-spin currents. Note also that for higher spin operators $\Delta = s+1$ in the free theory, so we expect the $M/N$ anomalous dimension to be non-negative for all values of $\lambda$. Calculations for higher-spin operators would be similar to those sketched here, but with more complicated vertex factors; these seem most feasible in our simplifying limit of $\lambda_N=0$.

As Figure \ref{plota} illustrates, the theory appears to contain interesting physics at first order in $M/N$ even when $\lambda_N=0$. We would like to perform a systematic study of the two-point and three point functions in this limit, particularly the two-point function of the stress tensor. It also would be interesting calculate the anomalous dimension of the scalar at higher orders in $M/N$, where the qualitative differences between a $1/N$ expansion and $M/N$ expansion are more apparent.

An interesting feature of feature of bifundamental theories in 3 dimensions is the novel Higgs mechanism \cite{Mukhi:2008ux}, studied in the more general context of non-supersymmetric Chern-Simons theories in \cite{Mukhi:2011jp}, which played a crucial role in the interpretation of \cite{Bagger:2007jr}. While \cite{Mukhi:2011jp} considers bifundamental theories coupled to scalars, it would be interesting to see how it might be applicable to the purely fermionic theory via say, a bosonization duality.

We also have not considered the $O(N)$, bosonic or critical versions of the theory. The bosonic theory has the additional complication of a classically marginal $\phi^6$ coupling, and requires a bit more care than the non-critical fermionic theory -- see, \cite{Banerjee:2013nca} for some interesting perturbative calculations in the bifundamental bosonic theory. We haven't considered the issues discussed in \cite{Bardeen:2014paa}, and these appear important to study. 

\subsection{Duality and M-theory limit}
The ``M-theory'' limit of a fermionic vector model with $N_f=1$, at finite\footnote{Here, we refer the Chern-Simons level in terms of a Yangs-Mills regularized theory as $k_{YM}$, which differs from the level used in the paper by a shift $k_{d}=k_{YM} - N$, where $N$ is the gauge group rank.} $k_{YM}$ and large $N$ is believed to have a dual description, via the bosonization duality proposed in \cite{Aharony:2012nh}, as a critical bosonic theory at finite\footnote{Though the duality is untested for finite $N$, based on its formal similarity to the duality in \cite{Giveon:2008zn} we might expect it to hold (with the addition of shifts in the Chern-Simons level $\pm 1/2$ \cite{Aharony:2012nh}.)} $N$ and large $k_{YM}$.  The duality seems to indicate that the ``M-theory'' limit exists for a non-supersymmetric Chern-Simons vector model, although gravitational behaviour is not expected.

In our simplifying limit, $\lambda_N=0$, the bifundamental theory is a $U(M)$ vector model with a very large number $N_f=N \gg M$ of flavors. In the alternative limit, $\lambda_M=0$, the theory is a $U(N)$ vector model with  $N_f=M \ll N$ flavors, whose $M/N$ corrections we are interested in. 

A conjectured generalization of the bosonization duality to $N_f$ flavors is presented in \cite{Aharony:2012nh}, which states the $U(M^F)$ level $k_{YM}^F$ theory coupled to $N_f$ flavors of fundamental fermions is equivalent to a critical $U(M^B)$ level $k_{YM}^B$ Chern-Simons theory coupled to $N^B_f$ flavors of fundamental bosons (with a parity transformation, see \cite{GurAri:2012is}):
\begin{eqnarray}
 N^F_f & = & N^B_f \\
 M^F & = & k_{YM}^B \\
 k_{YM}^F & = & M^B - N^B_f/2
\end{eqnarray}

This duality appears well tested if $N_f$ is small, and the gauge
group rank is very large; however, it has not been tested at $N_f/N$, and may not be expected to hold in the limit of $N_f$ is much larger than the gauge group rank $N$. In terms of $\alpha$ (which is $N_f/N$) and $\lambda=\frac{k}{N}$ the proposed duality takes the form:
\begin{eqnarray}
 1/\alpha^F & = & \lambda_{YM}^B\\
 1/\lambda_{YM}^F + 1/2 & = &  \alpha^B.
\end{eqnarray}

In terms of dimensional-reduction regularized $k$, the bosonization duality for the vector model states that critical $U(N)_k$ Chern-Simons theory with fundamental bosons is dual to a $U(k-N)_{-k}$ theory of fundamental fermions, where $|k| > N$. The supersymmetric ABJ theory \cite{Aharony:2008gk} is subject to a similar duality relating the $U(L+N)_k \times U(L)_{-k}$ theory to the $U(L)_k \times U(L+k-N)_{-k}$ theory, where $|k| > N$. If this generalizes to, say, a duality relating a critical $U(L+N)_k \times U(L)_{-k}$ bifundamental bosonic theory  to a $U(L)_k \times U(L+k-N)_{-k}$ bifundamental fermionic theory, (perhaps with $O(L)$ half-integer shifts in the Chern-Simons level(s) due to the fermions) it would be testable in an expansion in $\alpha=L/(L+N)$ -- though, again, the calculation would be tedious.

\subsection{Additional Comments}
Assuming the non-supersymmetric bifundamental theories considered here do have holographic duals; one would expect them to be some non-supersymmetric string theory vacua with a small negative cosmological constant, perhaps related to \cite{Kachru:2003aw}. We also note that \cite{Anninos:2014hia} appeared recently, which considers Chern-Simons theories coupled to ghost-like matter. These theories are non-unitary and expected to be dual to Vasiliev theories in de Sitter space. Similar bifundamental theories based on ghost-like matter could be constructed and may be dual to non-supersymmetric quantum theories of gravity in de Sitter space with a small positive cosmological constant.

Gravitational theories with and without dilatons/axions have been used extensively as models for strongly interacting quantum critical points, see e.g., \cite{Herzog:2007ij, Hartnoll:2007ih, Gubser:2008wz, Hartnoll:2009sz} for study of gravitational systems without a dilaton or axion. In, e.g., \cite{Goldstein:2009cv, Charmousis:2010zz, Goldstein:2010aw, Bayntun:2010nx}, the transport properties for black branes in various gravitational theories with both a dilaton and axion were considered. It would be interesting to study these nonsupersymmetric bifundamental theories at finite chemical potential in an $M/N$ expansion, along the lines of \cite{Yokoyama:2012fa,Aharony:2012ns,Jain:2013py}, and compare their behaviour to gravitational calculations, although studies at finite temperature would involve considerations of holonomy. 

Theories based on a single Chern-Simons gauge field provide a rich class of parity-violating IR fixed points for condensed matter systems \cite{Frohlich:1991wb}; bifundamental theories based on two Chern-Simons gauge fields with equal and opposite level provide a similarly rich class of parity-preserving IR-fixed points. The existence of highly supersymmetric bifundamental theories provided the resolution to some long-standing questions regarding the IR limits of 2+1 dimensional super Yang-Mills theory \cite{Schwarz:2004yj, Bagger:2007jr, VanRaamsdonk:2008ft, Aharony:2008ug}.  Analogously, the non-supersymmetric bifundamental theories mentioned here, which may also have a gravitational description in the large-$N$ limit, are natural IR limits for parity-invariant effective field theories arising in 2+1 dimensional condensed matter models, such as, algebraic spin liquids \cite{PhysRevB.66.144501} and variants thereof.

We hope to return to these issues in the future.

\section{Acknowledgements}
SP would like to thank Shiraz Minwalla for comments on a draft of this paper, and for introducing him to the idea of an $M/N$ expansion quite some time ago. SP would also like to thank his other collaborators in \cite{Giombi:2011kc}, Simone Giombi,  Sandip Trivedi, Spenta Wadia and Xi Yin, as well as Sachin Jain, V. Umesh and Shuichi Yokoyama for various discussions on related topics over the past few years. We would also like to thank Akshay Bedhotiya for discussions on some related calculations. Finally, we would like to thank the people of India for supporting research in string theory and the basic sciences.

\bibliography{CSBib}{}
\bibliographystyle{JHEP}
\end{document}